# Building AI Innovation Labs together with Companies[1]

Jens Heidrich, Andreas Jedlitschka, Adam Trendowicz, Anna Maria Vollmer

Fraunhofer IESE, Fraunhofer-Platz 1, 67663 Kaiserslautern, Germany
{ Jens.Heidrich | Andreas.Jedlitschka | Adam.Trendowicz | Anna-Maria.Vollmer }@iese.fraunhofer.de

**Abstract.** In the future, most companies will be confronted with the topic of Artificial Intelligence (AI) and will have to decide on their strategy in this regards. Currently, a lot of companies are thinking about whether and how AI and the usage of data will impact their business model and what potential use cases could look like. One of the biggest challenges lies in coming up with innovative solution ideas with a clear business value. This requires business competencies on the one hand and technical competencies in AI and data analytics on the other hand. In this article, we present the concept of AI innovation labs and demonstrate a comprehensive framework, from coming up with the right ideas to incrementally implementing and evaluating them regarding their business value and their feasibility based on a company's capabilities. The concept is the result of nine years of working on data-driven innovations with companies from various domains. Furthermore, we share some lessons learned from its practical applications. Even though a lot of technical publications can be found in the literature regarding the development of AI models and many consultancy companies provide corresponding services for building AI innovations, we found very few publications sharing details about what an end-to-end framework could look like.

## 1 Motivation

The topic of Artificial Intelligence (AI) is everywhere nowadays. Many studies and market research institutes highlight the importance and growing economic impact of AI in the future. According to [1], it is seen as an engine of productivity and economic growth. It may increase the efficiency of processes and stimulate the creation of new products, services, and business models. Accenture predicts that annual economic growth rates may double by 2035 [2]. PwC estimates that global GDP may increase by 14% by 2030 [3]. The McKinsey Global Institute says that 70% of companies will use AI technologies by 2030 [4]. In summary, all studies agree that there is a huge potential that AI can foster innovation and well-being if risks such as disruptive effects on the economy and society can be managed. Currently, legislators are about to define the general conditions of using AI systems [5] as well as ethics guidelines for trustworthy AI [6].

**Opportunities.** Based on our experience from AI-related projects with various companies, combinations of the following aspects drive them to think about using AI:

- *Operational excellence:* AI supports companies in increasing the effectiveness and efficiency of core processes. For instance, preventive maintenance of machines in a factory allows detecting issues with production machines early on and planning maintenance intervals. Image analysis allows finding defective parts more effectively. Typically, an increase in operational excellence results in saving costs and increasing revenues.

---

[1] To appear in: A. Schmietendorf (Publisher), "ESAPI light 2021." In: Berliner Schriften zu modernen Integrationsarchitekturen, Shaker-Verlag, Düren, April 2022, Vol. 27, ISBN 978-3-8440-8326-2.

- *Innovation:* AI supports companies in coming up with new innovative products and services that were not possible before. For instance, AI allows building highly automated and even autonomous machines, such as Collaborative Robots (Cobots), which can be used flexibly for different tasks in close collaboration with humans. Typically, such innovation leads to new business models and to new customer groups being addressed.
- *Customer intimacy:* AI supports companies in better understanding their customers (consumers as well as other businesses). For instance, AI allows analyzing the buying habits of customers or their general interests, and creating custom-tailored offers for them. Typically, this leads to more goal-oriented investments into product development and marketing and increases sales and revenues.

All three aspects might be good reasons to start thinking about the impact of AI on a company and about building up competencies and an AI strategy. However, there is a fourth aspect that should not be underestimated from the company perspective: competition. Everybody is talking about AI and its opportunities. There is huge pressure on companies who fear that their traditional markets and customers will be disputed by competing companies following a more AI or data-driven business model and thereby delivering the same or even better quality of products and services at lower cost or having more direct access to potential customers.

**Background.** AI itself is defined quite heterogeneously in the literature. We use a combination of the AI definitions according to Marvin Minsky [7], one of the founders of AI concepts, and according to Nils J. Nilsson [8]. Thus, we define AI as "the science of making machines do things that would require intelligence if done by men", and "… intelligence is that quality that enables an entity to function appropriately and with foresight in its environment." In addition, in the literature, the term AI may also be used as an abbreviation referring to AI methods, AI components, AI technologies, or AI systems. However, in the context of this article, we use it as an umbrella term including the theoretical background as well as practical implementations.

There are many different approaches aimed at making machines do things that would require intelligence if done by men. For instance, at the beginning of AI research in the 1960s, symbolic AI [9], which simulates high-level conscious reasoning, was used in attempts to create machines with general intelligence. Nowadays, highly data-based AI approaches, which are largely directed at specific problems with specific goals rather than general intelligence are used. The reason for this is the general availability of data on the one hand and the technologies available for analyzing this huge amount of data within a reasonable amount of time on the other hand. One approach in this regard is Machine Learning (ML) or, more specifically, Deep Learning (DL). The resulting model (e.g., an Artificial Neural Network) is generally not comprehensible to humans due to its structure and inherent complexity, and thus the decisions made by DL systems are often difficult to understand. However, in practice, simpler models, such as statistical classifications (e.g., Random Forest) may often be sufficient and result in far less complex and more transparent models. When we refer to AI approaches in this article, we refer to all data-based approaches.

**Challenges.** The reasons for failed AI projects are manifold [10]. Often, failure is related to being aligned neither with the strategic objectives nor with the operational capabilities of a specific organization [11]. When companies want to come up with a strategy for dealing with AI, they are typically faced with the following challenges (based on projects we did with customers from different industries):

- *Unclear use case:* One of the biggest challenges is finding the right AI use case. If the use case is unclear or does not return any value for the organization and return on its investments, AI projects are doomed to fail [10]. Companies are tempted to deal with the technologies and infrastructure first, and think about the use case second. In fact, it should be the other way around. Use cases should be driven by clear business needs and must fit in with the organization's capabilities.

- *Availability of data:* Often, there is a lack of data for building AI models or the available data has some deficiencies that make it difficult to analyze, or it is simply the wrong data. For example, it has been found that 90% of deployed data lakes end up being useless as they are overwhelmed with information assets captured for uncertain use cases [11]. Furthermore, mapping data from different sources (e.g., data from two different sensors) is often challenging (e.g., because time stamps are not synchronized perfectly). Therefore, it is important to rapidly try out potential use cases, build a proof of concept based on available data, and determine what results can be obtained realistically from the available data, or what investments could be made into collecting new data or improving data quality.
- *From proof of concept to product/service:* Once a use case has been successfully found and promising AI models have been built from the available data, building a product or service based on this can also be challenging. AI models often have to be integrated into an existing product or service that can be deployed to potential customers. This does not only require data science expertise, but also fundamental expertise in software and systems engineering and strong collaboration between people having these competencies.
- *Maintenance of AI systems:* If an AI system is used by customers, its performance should be monitored. For high-risk AI systems used in critical application cases (such as autonomous driving functions), this is demanded by law. However, even for less critical systems, this makes sense as it is necessary to monitor the quality of the AI model, detect performance issues, and adapt or learn and deploy new models if required. This requires companies to have the right structures and processes in place for DevOps [12] in order to be able to react quickly to issues and deploy fixes or new requested features on-the-fly, e.g., via updates over the air (OTA).
- *Qualification of employees:* Last, but not least, there is a lack of necessary knowledge and experience in companies when it comes to AI and often also when it comes to software engineering competencies. Large organizations frequently have their established data science and development teams, but for smaller and medium-sized companies not coming from the software domain, this is often challenging. Some of these gaps can be compensated for by working with external partners. However, depending on the criticality of the intellectual property that goes into an AI model, it is often not an option to outsource these critical parts. Competencies must also be built up internally by hiring new people or qualifying existing personnel.

All these challenges are even more critical in the context of collaborative enterprises. In digital ecosystems or digital platforms, data and AI-driven business models and technical solutions must be aligned with both the individual and the collective interests of multiple organizations. According to [13], data and ecosystems are key drivers of future trends in software engineering.

**Contents.** In this article, we want to demonstrate how to address the challenges shown above by introducing a framework called *AI innovation labs* in companies. This framework supports companies in coming up with and implementing AI use cases that add real business value instead of them starting to invest into infrastructures and tools without a clear purpose. In the remainder of this article, we will first describe the framework, then illustrate lessons learned from its practical applications, discuss related work, and finally conclude with a brief summary and an outlook on future work.

## 2 AI Innovation Labs

An *AI innovation lab* is a framework used for systematically creating, implementing, and evaluating AI use cases for a company that have a clear business impact. It supports the whole process, from creating ideas via prototyping to rolling them out as part of products and services. It supports companies in systematically answering the following key questions:

- How to benefit from Artificial Intelligence? What is the added value? What is the impact on the business model and the customers?

- What are meaningful AI use cases? Is the required data available and does it have a sufficient level of quality?
- What competencies and infrastructures are needed and are they available?
- What investments are necessary for deploying an Artificial Intelligence solution?

In the following sections, the general idea and framework of *AI innovation labs* will be described. First, we will present the general approach for coming up with AI innovation ideas. Second, the innovation process for creating, implementing, and evaluating those ideas will be presented. The remaining three subsections will go into the details of the conceptual components used in the framework: the definition of business solutions, the implantation of technical solutions, and how these solutions are evaluated using a staged labs and prototyping approach.

## 2.1 AI Innovation Ideas

As of today, two main strategies are being followed to exploit data for creating business value: the data-driven strategy and the business-driven strategy (see Figure 1). Although their starting points are very different, both aim at finding innovative ideas for solving business problems or exploiting new business opportunities.

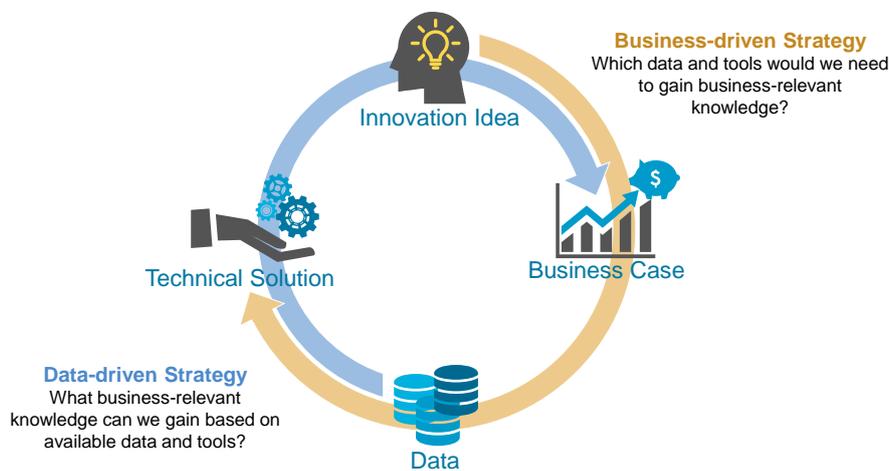

*Figure 1: Business- and data-driven AI strategies*

The *data-driven strategy* takes whatever data is available and explores it using AI or data-based technical solutions built in an ad-hoc manner to gain business-relevant knowledge. The *business-driven strategy* starts with the business needs and derives potential business cases in order to come up with appropriate technical solutions for gaining relevant business knowledge. Both strategies have their strengths and weaknesses (see Table 1) and should thus be considered as complementary rather than competitive in practice.

One way of combining both strategies in practice is to involve both domain expertise and the results of a data analysis in the derivation of potential AI business cases. For example, representatives of business experts and analysts can search for innovative ideas during a joint creativity workshop. The business experts can contribute their knowledge and understanding of business, whereas the analysts can share their knowledge of available data and existing analysis results. Expert knowledge can additionally be combined with analyses of already available data – either already existing analyses or simple analyses performed in an ad-hoc manner. Innovation ideas can then be concretized in the form of business cases.

A *business case* defines the path towards an AI or data-driven business model and provides arguments for implementing this path by comparing it to alternative paths, so-called solution options, that are not based on AI. Comparison is done systematically based on detailed information including:

- *Assumptions* upon which the solution's ability to achieve the business outcome is based
- *Benefits* expected after implementing the solution

- *Costs* estimated for implementing the solution
- *Time* until real benefits are gained from the solution
- *Risks* associated with the implementation and deployment of the solution

In order to get a chance of being deployed, the selected solution must demonstrate a cost-benefit ratio that is significantly better than that of alternative solutions.

*Table 1: Comparison of business-driven and data-driven AI strategies (+ = strength, o = neutral, - = weakness)*

| **Business-driven Strategy** | **Data-driven Strategy** |
|---|---|
| *(+) Business-oriented:* Solutions are driven by business goals and needs and fit the capabilities and constraints of a specific organization. | *(+) Data-oriented:* Solutions are driven by the data to be analyzed and explore insights from the data that could lead to business value. |
| *(-) Tunnel vision:* Solutions are more likely to be limited to the known business context and less likely to explore completely new fields. | *(o) Open mind:* Explorative data analysis has the advantage that it may lead to disruptive insights. However, ideas are limited to available data. |
| *(+) Systematic:* Solutions are based on well-founded procedures and selections of alternative options and implementation decisions. | *(-) Unsystematic:* Data is explored using familiar techniques in a trial-and-error process guided by experiences and available resources. |
| *(+) Foreseeable cost-benefit ratio:* Potential benefits and costs of implementing a specific solution are assessed in advance. | *(-) Unforeseeable cost-benefit ratio:* Benefits can only be estimated retrospectively after the potential business value becomes clear. |
| *(+) Low risk:* Solutions are aligned with business needs and operational capabilities in advance. This reduces the risk of failure. | *(-) High risk:* The business value of explorative analyses can range from disruptive insights to no insights at all. This increases the risk of failure. |
| *(-) High upfront effort:* Deriving and evaluating AI use cases beforehand requires more effort upfront before implementing a solution. | *(+) Low upfront effort:* Solutions are tried out in an ad-hoc fashion and may lead to relevant AI use cases. Less upfront effort is required. |

## 2.2 AI Innovation Process

Based on the business case, an appropriate prototype for a technical solution can be constructed. This includes identifying the necessary data, AI technologies, and infrastructure while considering organizational capabilities and constraints such as available financial and human resources or already available infrastructure and competencies. Data and AI-driven business innovation is an iterative process (see Figure 2) where innovation ideas and the associated technical solutions are evaluated and matured. In each iteration, a viable prototype of the innovation ideas and technical solutions is designed and evaluated against success criteria such as business profitability or technical feasibility. Based on the feedback, the innovation ideas and technical solutions are improved iteratively. The main objective of these improvement cycles is to learn about the underlying problem and to explore alternative solutions. That is, the goal is to sort out and reject unpromising ideas as early as possible and to continue focusing on the promising ones.

In each iteration of the innovation cycle, the business potential of a specific solution is assessed regarding the expected business value and the cost it will require to be implemented. This *AI innovation process* supports the derivation, evaluation, and maturation of AI-driven innovations to minimize risk and losses caused by investing into the wrong solutions.

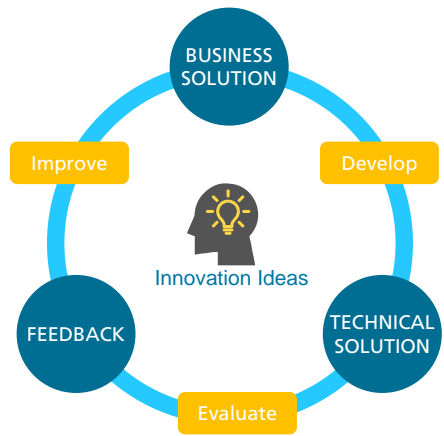

*Figure 2: Business innovation process based on the Lean Startup [14]*

Figure 3 presents an overview of the conceptual components used in our AI innovation process. All components are centered around the *business case*, which defines the path towards an AI-driven business model and integrates the outcomes of the surrounding building blocks.

- *Scope and current situation:* First, the organizational scope and the current situation must be understood. This typically includes sources of potential business challenges and opportunities and factors influencing the feasibility of potential technical solutions as well as the current business model pursued.
- *Business solutions:* Based on the current situation, innovation ideas are created. In essence, AI or data-driven business solutions are identified that improve the existing business model or lead to new business models. Each business solution is subjected to an evaluation regarding its chances of success and organizational readiness to realize it. Based on the outcomes of the evaluation, potential business solutions are selected and improved or abandoned. The business solutions with the highest chances of success are summarized in the form of business cases. A business case not only documents specific business needs and the proposed business solution, but also motivates its implementation as a technical solution by demonstrating its advantage compared to alternative solutions.
- *Technical solutions:* Only after specific business solutions have been evaluated positively does the innovation process continue with the development of corresponding technical solutions (based on AI or data). Technical solutions are developed in an iterative process. With each cycle of this process, the technical implementation becomes more mature: from a pure paper-based concept to an initial proof of concept (e.g., regarding data and model quality [15]) to a solution that is ready to be deployed and rolled out to products and services. After the end of each cycle, the readiness of the organization to implement and deploy a specific technical solution is assessed and the gap between the required and available organizational capabilities is determined. A potential capability gap can be addressed by either adjusting the business solution or the associated technical solution (e.g., in terms of necessary data, analysis methods, and infrastructure).

The motivation for developing the business solution first is that it is typically significantly cheaper to test and (potentially) fail here than to fail with a technical solution for which prototyping and testing already require considerable investments into infrastructure and personnel. Valid business solutions are mainly developed as part of an initial potential analysis together with the paper-based concept for corresponding technical solutions. In a series of labs and prototyping stages, the technical concept is refined and implemented based on real data. If successful, the resulting technical solution is ready to be deployed to a product or service. If not successful, it is either rejected or reworked based on the evaluation results after each iteration.

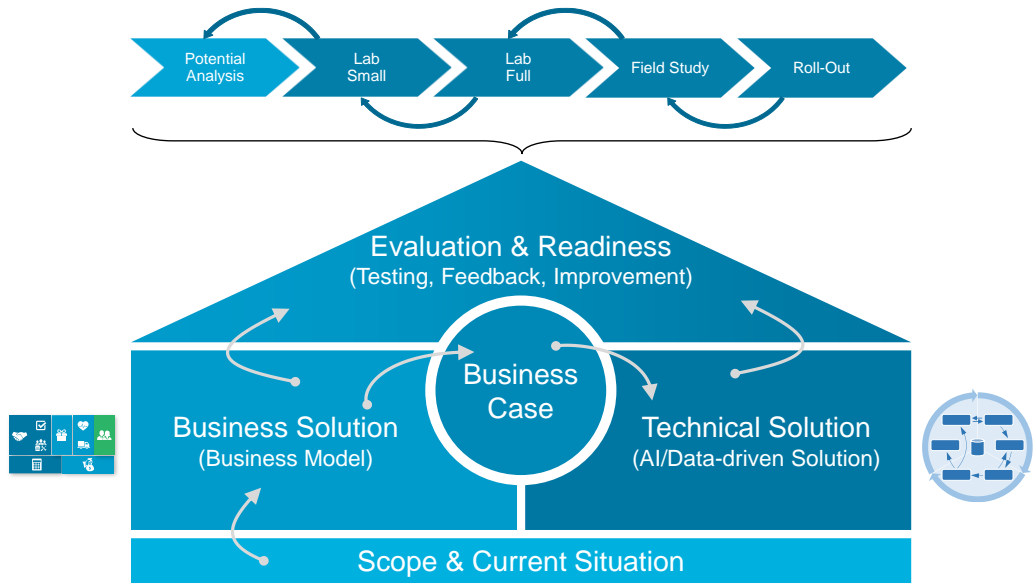

*Figure 3: Conceptual components used in the AI innovation process*

## 2.3 Definition of Business Solutions

In practice, the ultimate goal of employing AI is business innovation. We aim to use AI to improve an existing business model or create a new business model. The success of an AI-driven business model is measured by the value it delivers and the cost it requires to be realized, including the implementation of a specific technical solution. To be successful, the use of AI must create significantly more value than costs. As discussed at the beginning, there are different opportunities in using AI, such as improving operational excellence, coming up with new innovative products and services, or getting to know one's customers better (see Figure 4).

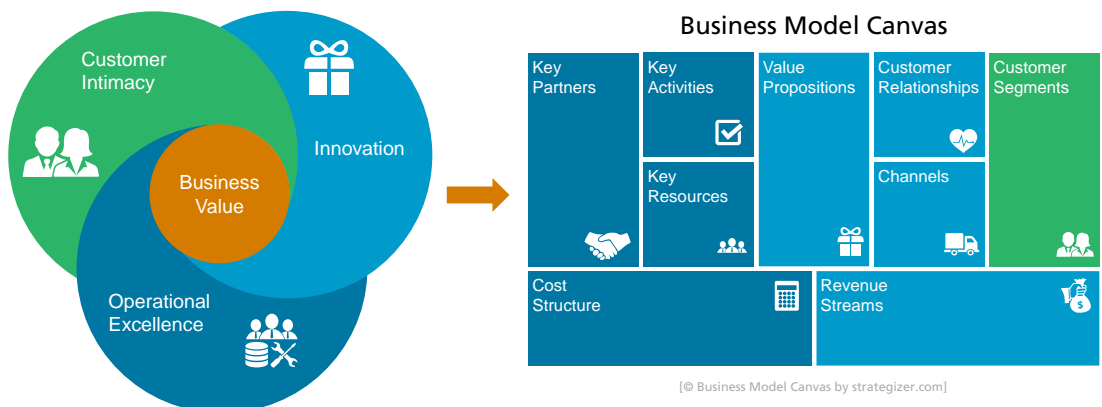

*Figure 4: Business value creation areas of a business model*

In practice, AI or data-driven business solutions are typically found using creativity techniques in a series of workshops. At the very beginning of the process, current trends and impact factors fitting the scope and current situation of a company are presented and discussed. Based on this input, potential (old and new) customer groups are identified and pains and gains of these groups are carefully analyzed. A systematic approach for documenting this is called Value Proposition Canvas [16].

When the pains and gains are known, business solution ideas are brainstormed using creativity techniques (such as brain storming). The ideas are typically grouped into clusters and then prioritized based on predefined criteria (such as innovation degree or customer interest). Those ideas with the highest priority are selected to be refined. For this purpose, the implications of the selected solution ideas on the existing business model of a company are analyzed. This may lead to a completely new business model based on the AI or data-driven solution idea or to the existing business model being adapted accordingly.

For the description of business models, we typically make use of the Business Modeling Canvas (BMC) approach [17]. Figure 4 illustrates the different sections of a BMC and how the different value creation areas affect those sections of a business model. For instance, product or service innovations could lead to an improved value proposition, allow for new revenue streams, new channels to customers, or improved customer relationships.

Typically, different groups participate in such creativity workshops: people who understand the business of the company (such as mid or high-level management and business developers), people with an understanding of the current technical solutions and data available in the company (such as CTO, IT or software people), and finally people who contribute AI or other data competencies (such as internal or external data scientists). At the end, the success depends on having mixed groups working together bringing in different perspectives. Depending on the number of workshop participants, different groups may also compete with each other for the "best" AI or data-driven business solution. For that purpose, the solution ideas may further be illustrated by preparing an initial fake product or service name, selling pitches to customers, or doing product presentations.

All elaborated business solutions (business models) are finally evaluated regarding different aspects, including their plausibility, profitability (at least roughly), and expected desire of potential customers. A successfully evaluated business solution is documented in a business case, which not only includes the previous business model and the new/adapted AI or data-driven business model, but also the rationale for pursuing this model, including alternative options that were evaluated in the process. Based on the current capabilities of the company, a plan is defined for implementing and evaluating a corresponding technical solution.

### 2.4 Implementation of Technical Solutions

The technical solution is developed based on a business case in an iterative manner. Typically, it starts with a theoretical / purely paper-based technical solution concept, which is then implemented and tried out in different iterations until its maturity is sufficient to be rolled out as part of a product or service.

In this regard, we make use of the CRISP-DM process, which stands for Cross-Industry Standard Process for Data Mining [18] (see Figure 5). It describes six steps that are essential for data mining and any kind of data analytics and is widely used in research and industry:

- *Business Understanding:* The goal of this step is to understand the business objectives of the organization, define analysis goals, and set up a project plan. As part of this step, the target usage scenario for AI is defined based on the description given in the associated business case. Furthermore, the target scenario stakeholders and their information needs are specified.
- *Data Understanding:* The goal of this step is to collect, characterize, und understand the data and its meaning and analyze the quality of the available data. Initially, this includes the identification of data sources in the company and a characterization of the available data as well as the data required for the target usage scenario. In later iterations, this includes accessing and retrieving data from those data sources and analyzing its quality. If required data is not available or of bad quality, an attempt is made to access more data or improve the quality of existing data (e.g., by filtering or data healing). If this fails, the usage scenario is disregarded and the focus is shifted to a different business case.
- *Data Preparation:* The goal of this step is to select the data to be analyzed and to clean and format it. This goes hand-in-hand with analyzing the quality of data during data understanding. All known deficiencies of the data (e.g., quality or formatting issues) are treated in the best possible way and the data is prepared for model building.
- *Modeling:* The goal of this step is to choose the right technique for model building, create the model, and analyze the quality of the model. This includes trying out different solution ideas, that is, AI modeling techniques on the data (such as different Machine Learning algorithms) and evaluating the quality of the resulting models (such as precision or recall rates). During the concept stage, potential algorithms are preselected in this step, which are tried out on real data in later iterations.
- *Evaluation:* The goal of this step is to evaluate and review the results and determine follow-up steps. This step marks the final evaluation of one iteration. During this evaluation, the decision is made whether to continue working on the usage scenario, whether some adaptation of the business case is required, or whether the model has reached the quality required for deployment.
- *Deployment:* The goal of this step is to plan how to deploy and maintain the model in practice, and to document the outcomes and lessons learned. As part of this step, the system and software architecture into which the model is to be integrated is defined and set up. Furthermore, steps are taken for maintaining the model in the future. This could include, for instance, observing the performance of the model during runtime of the product or service, and improving / rebuilding the model in case of issues. In practice, this step can be seen as the touchpoint to a company's traditional software and systems development process.

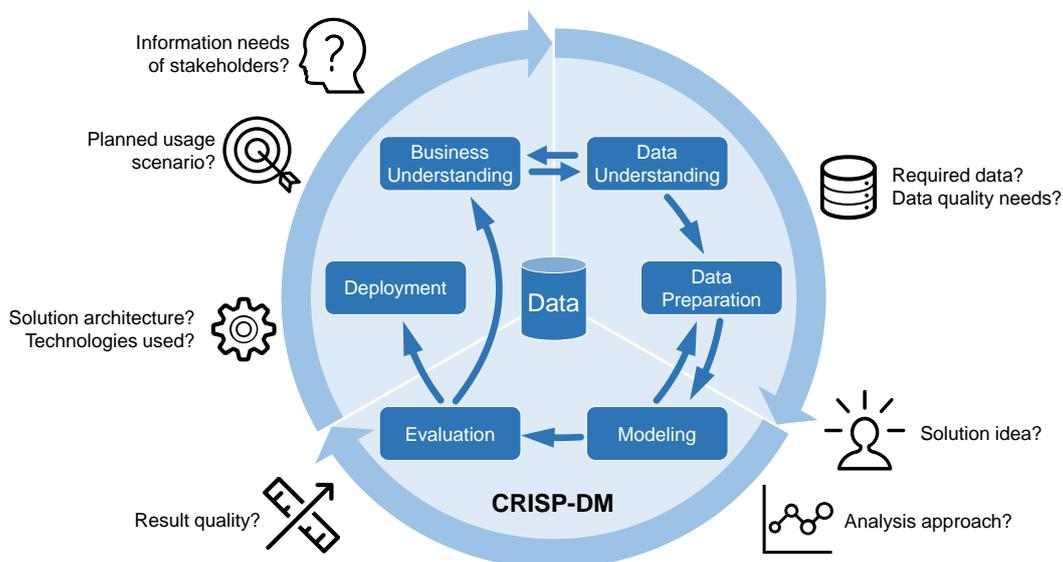

*Figure 5: Technical solution development based on CRISP-DM*

## 2.5 Staged Labs and Evaluation of Solutions

To minimize the risk of failing and reduce potential business losses, the business and technical solutions are evaluated and matured throughout several specific lab and piloting stages before they can be deployed in a productive environment (see Figure 6). After each step, the results are evaluated according to pre-defined criteria. Note that business solutions and corresponding technical solutions are strongly connected with each other. For instance, the accuracy of a prediction algorithm may directly influence the reliability of customer-relevant forecasts and thus affect customers' desire for the realized business solution.

At each stage, the technical solution concept, the associated business solution, and the organizational readiness are revised based upon the evaluation outcomes. In the very first stage, an initial technical solution concept is blueprinted and evaluated conceptually without any practical implementation. In the following stages, specific "in-use" aspects of the technical solution concept are verified after being implemented in test environments. For example, the performance and scalability of the selected data analysis technologies are evaluated in a lab environment using real or simulated data. Evaluating the integration with existing infrastructure and processes as well as user acceptance, on the other hand, requires piloting the technical solution in the intended target environment. Based on the outcomes of the implementation, the technical solution and the corresponding business solution are revised and re-assessed regarding business impact and organizational readiness. Only if the technical solution has successfully run through all intermediate stages is it rolled out into a productive environment.

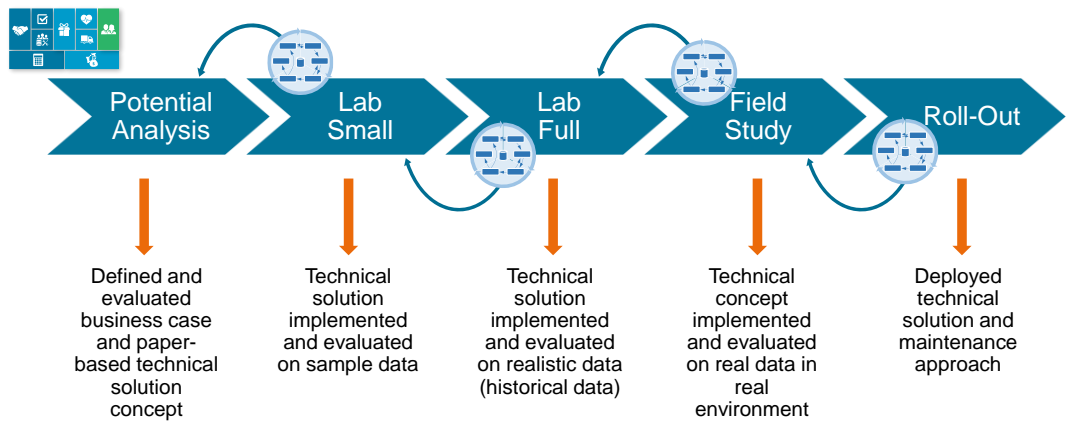

*Figure 6: Staged development of business and technical solutions*

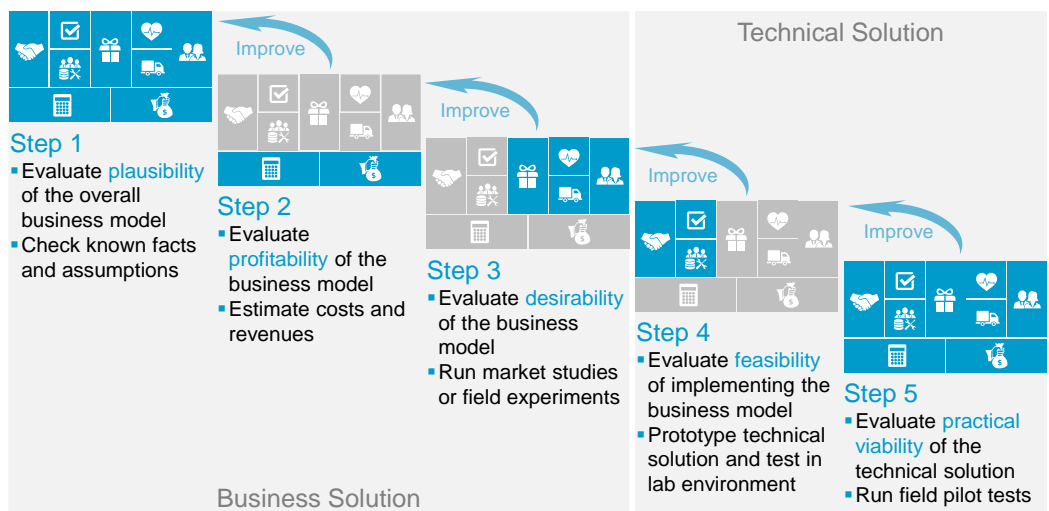

*Figure 7: Comprehensive evaluation of the business and technical solution*

Figure 7 gives an overview of the evaluation criteria for business and technical solutions. Each step shown refers to certain parts of a business model:

- Step 1 focuses on the overall *plausibility* of the model. Therefore, the different elements of the BMC are checked for a consistent story line. This includes, for instance, that (1) the defined value proposition fits the customer segments, channels, and relationship; (2) all key activities, resources, and partners for developing the value proposition are specified; and (3) cost and revenue streams are defined properly. Furthermore, it should be analyzed what parts of the BMC rely on known facts and where assumptions are made that must be clarified at a later stage (e.g., regarding value streams or cost).
- Step 2 focuses on evaluating the *profitability* of the business model. Therefore, the costs for developing the value proposition and the revenue streams from customers are estimated more thoroughly. To do so, various scenarios can be considered based on transparent assumptions regarding how the market will evolve and how many customers can be attracted.
- Step 3 focuses on evaluating the *desirability* of the business model from a customer's perspective. Therefore, the attractiveness of the value proposition is assessed for current and future customers. This could be done based on expert judgment or by running market studies or field experiments.
- Step 4 focuses on evaluating the *feasibility* of the technical solution that implements the business model. Therefore, prototypes of the technical solution are developed and tested in a lab environment. This provides feedback regarding the required key activities, resources, and partners.

- Step 5 focuses on evaluating the *practical viability* of the developed technical solution as well as the overall business model. Therefore, pilot tests are performed with the technical solution in the field.

# 3 Lessons Learned from Practical Applications

We applied the *AI innovation labs* framework in different flavors and development stages with various industry companies over the course of the last nine years in various application domains (e.g., [19]). It is the nature of innovation processes that the concrete solutions ideas that were created following the process contain a lot of intellectual property from the companies. Therefore, for confidentiality reasons, we cannot share details about the outcomes, but we summarize some lessons learned from the practical application of the approach.

- *Right culture and mixture of participants:* Even though the approach helps to systematically come up with AI or data-driven use cases and solution ideas that are aligned with the business models of a company, it largely depends on the participants of the workshops performed during the potential analysis how good or bad those ideas are. Having the right mixture of people, representing the business know-how, the technical know-how, and the AI know-how working together is key to coming up with good ideas. This also depends on the culture of a company and how open this culture is to bringing in new ideas and openly sharing thoughts with each other.
- *Data collection and preparation effort:* In recent years, some companies worked hard to build up data lakes, collecting all sorts of data across the organization. However, depending on the identified AI use case, the requirements and quality demands for data are quite different. In practice, we often see that the available data requires a significant amount of cleaning and preparation before it can be used or that investments into collecting high-quality data are required. The reason for this is often that the initial reason for collecting the original data was not to build data-driven models.
- *Integration with software and systems engineering:* In order to successfully deploy an AI solution, it is not sufficient to let a team of data scientists work on its implementation. Typically, an AI solution is part of a larger system or service and must be integrated into the software and systems architecture. This requires the data science team to cooperate closely with the software and systems engineering teams of a company. Involving these people in the endeavor right from the beginning is key to its success.
- *Fail early:* When developing AI solutions, it is important to notice that some solution ideas may lead to success and some to failure. In our experience, it does not make sense to treat the process in a waterfall-like manner and spend most of the time thinking about the perfect idea. Most innovation processes rely more on quickly trying ideas out in an agile, prototyping-oriented manner. Because if an idea is condemned to fail, we want to know this as early as possible in order not to spend any further effort and money on it, but redirect these to other ideas.
- *DevOps:* Typically, an AI model requires maintenance over time. Its performance is likely to decrease if the real data onto which it is applied changes compared to the training and test data used for building the model. Therefore, it makes sense to start thinking about mechanisms for updating the model right from the start and to establish a DevOps (also called MLOps specifically in Machine Learning) approach in which the development and operation of the AI solution is brought together by feedback cycles and data regarding its performance in the real world. This revision includes the technical as well as the business solution, which might both be impacted by the performance of the AI model.

In the different incarnations of the *AI innovation lab* framework we created over time, we tried to address these aspects in the best possible way.

## 4 Related Work

Of course, the whole field of innovation research is not new and contains a rich set of approaches and techniques for different areas such as economics, sociology, and technology management [20]. When we started working on the *AI innovation lab* framework in 2013, there were only bits and pieces available, but we found no comprehensive process or concept in the literature for systematically designing and developing AI solutions. As stated above, our approach does not try to reinvent the wheel, but builds on well-known concepts from Lean Startup [14], creativity techniques [21], Value Proposition Canvas [16], Business Model Canvas [17], CRISP-DM [18], and DevOps [12]. It combines those existing concepts into an overall framework for designing and developing AI solutions. In the first years of using the framework, our focus was on innovations from Big Data [22].

Today, with the continuing rise in popularity of terms related to AI, digital transformation, Big Data, or data spaces, we find more approaches focusing specifically on creating innovations related to AI or data. Every larger consultancy company offers services around creating ideas and use cases in the context of the digital transformation (making more or less use of AI). We find consultancy companies offering their services across domains or focusing on specific application fields, such as Smart Production / Industry 4.0. However, we still believe that there is value in publishing more details in the literature about what a successful framework could look like and what can be learned from its application because only few details can be found about approaches that are similar to ours (such as the Value Co-Creation process by Siemens [23]).

In addition to approaches focusing on stimulating innovation, there are also maturity-oriented approaches for evaluating and developing data science competencies in an organization, such as the Data Science Maturity Model [24] or the Data Science Readiness Assessment [25]. Typically, those approaches either focus on the competencies of an organization as a whole or those of an individual within the organization. They do not address the development of a specific business case.

## 5 Conclusions

This article introduced the *AI innovation labs* framework, which aims to find an AI or data-driven technical solution for a specific business innovation idea that provides a clear return on investment. That is, it comes up with a technical solution that delivers the best trade-off between the potential business value and the investments required for implementing and deploying it. The development of a suitable solution may require several iterations in which both the anticipated business solution and the required technical solution are evaluated. From our perspective, this approach has several benefits:

- *Creativity techniques* applied together with all relevant stakeholders (such as domain experts and data scientists) help to come up with *innovative AI or data-driven solution ideas*.
- The *innovation process* helps to provide a clear link between AI or data-driven technical solutions and their actual business value and a *transparent decision-making process*.
- *Structured templates* help to systematically keep track of all discussed business and technical solutions and to have *clear rationales* for discarding or supporting them.
- *Evaluation criteria* help to focus on *plausible* business solutions that are *profitable* and *desired* by customers, and on technical solutions that are actually *feasible* and *viable*.
- The highly *iterative approach* guarantees early feedback on solution ideas and *prevents wrong investments* into technical solutions.
- The specific *lab and piloting stages* help to determine the *readiness of an organization* to pursue the desired business cases and the gap where it needs to invest into competencies and infrastructure.

The focus of current and future research is on the application of the *AI innovation labs* framework together with more industrial partners from different domains. We plan to analyze patterns for typical business and technical solutions, which could provide good stimuli from which company-specific solutions could be derived. Furthermore, we want to develop a transfer strategy for *AI innovation labs* using train-the-trainer concepts for creating multiplicators in companies. Right now, *AI innovation labs* are largely driven by us as applied researchers. In the future, we want to act as a transfer partner enabling companies, and especially SMEs, to build up and operate their own innovation labs. This will hopefully lead to more self-reliance of companies in coming up with AI or data-driven solutions and allow them to compete with companies newly entering their established markets with an AI or data-driven business model.